\documentclass[aps,prb,reprint]{revtex4-1}

\usepackage[utf8x]{inputenc}
\usepackage[english]{babel}
\usepackage{amsmath}
\usepackage{amsfonts}
\usepackage{verbatim}
\usepackage{color}
\usepackage{hyperref}
\usepackage{xspace}
\usepackage{grffile}
\usepackage{todonotes}
\usepackage{multirow}
\usepackage{xr}
\usepackage{comment}

\begin{document}

\title{Convergence and pitfalls of density functional perturbation theory phonons calculations from a high-throughput perspective}

\author{Guido Petretto}
\affiliation{Institute of Condensed Matter and Nanoscience (IMCN), Universit\'{e} catholique de Louvain, B-1348 Louvain-la-neuve, Belgium}

\author{Xavier Gonze}
\affiliation{Institute of Condensed Matter and Nanoscience (IMCN), Universit\'{e} catholique de Louvain, B-1348 Louvain-la-neuve, Belgium}

\author{Geoffroy Hautier}
\affiliation{Institute of Condensed Matter and Nanoscience (IMCN), Universit\'{e} catholique de Louvain, B-1348 Louvain-la-neuve, Belgium}

\author{Gian-Marco Rignanese}
\affiliation{Institute of Condensed Matter and Nanoscience (IMCN), Universit\'{e} catholique de Louvain, B-1348 Louvain-la-neuve, Belgium}

\begin{abstract}
The diffusion of large databases collecting different kind of material properties from high-throughput density functional theory calculations has opened new paths in the study of materials science thanks to data mining and machine learning techniques. Phonon calculations have already been employed successfully to predict materials properties and interpret experimental data, e.g. phase stability, ferroelectricity and Raman spectra, so their availability for a large set of materials will further increase the analytical and predictive power at hand. Moving to a larger scale with density functional perturbation calculations, however, requires the presence of a robust framework to handle this challenging task. In light of this, we automatized the phonon calculation and applied the result to the analysis  of the convergence trends for several materials. This allowed to identify and tackle some common problems emerging in this kind of simulations and to lay out the basis to obtain reliable phonon band structures from high-throughput calculations, as well as optimizing the approach to standard phonon simulations.
\end{abstract}

\maketitle

\section{Introduction}

In the latest years, the development of efficient density functional theory (DFT) software packages and the constant increase of the computational capabilities of modern supercomputers have opened new possibilities in the field of material science, including the discovery of new materials based on the high-throughput screening of large number of compounds\cite{curtarolo2013high}. Following this trend, several open databases have been created containing a number of properties obtained from DFT calculations \cite{Jain2013,Curtarolo2012227,Saal2013,nomad,OpenMaterialsDatabase} thanks to \textit{ad hoc} frameworks developed to help the automation of the whole process\cite{fireworks2015,Ong2013314,Curtarolo2012218,Pizzi2016218,ASE}. High-throughput techniques have not only allowed to search for materials candidates for a specific application but also to use data mining and machine learning techniques to understand trends and limitations in materials science \cite{Jain2016,varley2017,Chen2016,isayev2017universal,ward2016,NATH201682}.

Building on the experience gained so far and on the tools made available, it is now possible to address the calculation of more and more involved quantities for a vast number of materials, with  efforts that are ongoing to analyze properties obtained with calculations beyond the ground state \cite{Legrain2017,phdb,Petousis2016,dejong2015,vansettenGW,plata2016}. Vibrational properties in solids are essential to many applications as they govern several materials properties (e.g. vibrational entropy\cite{Karki2000,Togo2010}, thermal conductivity\cite{Seko2015,Lindsay2014,Romero2015}, ferroelectric and ferroelastic transitions\cite{Bousquet2010,Togo2008}). Phonon computations can be performed very efficiently using the Density functional perturbation theory (DFPT)\cite{BaroniDFPT2001} providing a computational access to these important properties.

Before moving to large scale calculations it is mandatory to implement a robust procedure to handle the whole process and provide a reliable set of input parameters that suits most of the possible cases that will be considered. In this regard, it has been shown how the validation of the results is an element of great relevance when approaching the high-throughput regime\cite{Jain20112295,Petousis2016}. 

In this paper we present the outcome of a study concerning the convergence rate of phonon related quantities with respect to the sampling of the Brillouin zone alongside a validation of the results generated within an automated process. In addition, since performing a considerable amount of simulations provides a stress test for the DFT and DFPT code, we have been able to identify some subtleties and pitfalls, that tend to show up under certain conditions and that may go unnoticed in a standard approach. Given the substantial difference in terms of requirements between metals and semiconductors and the difficulty of expressing a general recipe that would allow to capture peculiar phenomena as the Kohn anomalies, we have limited our analysis to the case of semiconducting materials.

In the following, after the definition of the theoretical framework (Section~\ref{sec:method}), we will first discuss the evolution of the phonon frequencies with respect to the density of \textit{k}-points and \textit{q}-points in Sections~\ref{sec:kpt_conv} and \ref{sec:qpt_conv}, respectively, highlighting the problems related with the sampling. In the second part we proceed to demonstrate the reliability of our results through a validation process, involving also a comparison with experimental data (Section~\ref{sec:validation}).

\section{Formalism and methodologies}
\label{sec:method}
In this study we considered a set of 48 semiconducting materials. These were chosen to cover a variety of elements of the periodic table and all the crystal systems, with different system sizes and band gaps. Detailed information about the properties of each material can be found in the Supplemental Material (SM).

All the simulations were performed with the ABINIT software package\cite{abinit2005,abinit2009,abinit2016}, which relies on the DFPT formalism to carry on calculations related to phononic and electric field perturbations\cite{BaroniDFPT2001,GonzeDFPT1997a,GonzeDFPT1997b}. The Perdew-Burke-Ernzerhof (PBE) \cite{PBE1996} was used as exchange-correlation (xc) functional. While this standard approximation has been demonstrated to underperform for phonon frequencies compared to other exchange-correlation functionals, it still produces results in reasonable agreement with experimental data \cite{He2014} and we assume that our conclusions will still hold for other local and semilocal xc functional approximations (e.g., LDA and PBEsol).

Optimized norm-conserving pseudopotentials (ONCV)\cite{Hamannn2013} were used for all the elements treating semi-core states as valence electrons for transition metals (as available from the pseudopotentials table Pseudo-dojo version 0.2\cite{pseudodojo}). The cutoff was chosen independently for each material according to the values suggested in the Pseudo-dojo (see SM). These pseudopotentials and the cutoff values have been carefully tested with respect to all electron codes \cite{Lejaeghereaad3000} and with respect to the fulfillment of the acoustic sum rule (ASR). Checks on the ASR and the charge neutrality \cite{GonzeDFPT1997b}, that are sensitive to the cutoff, were also performed on each material and the convergence with respect to the cutoff was further verified for problematic cases.

The splitting between longitudinal and transverse optical mode (LO-TO) was taken into account through the calculation of the nonanalytic term containing the Born effective charges $Z^*$ (BECs)\cite{GonzeDFPT1997b}. Phonon frequencies at generic \textit{q}-points were obtained through Fourier interpolation and, in that case, the ASR and the charge neutrality at the $\Gamma$ point were enforced explicitly.

For all the materials, the unit cells were standardized according to Ref. \citenum{Setyawan2010299} and each was relaxed until all the forces on the atoms were below 10$^{-6}$ Ha/Bohr and the stresses were below 10$^{-4}$ Ha/Bohr$^3$.

All the calculations were carried out using Fireworks as a workflow manager\cite{fireworks2015} integrated with the different libraries: pymatgen\cite{Ong2013314}, abipy\cite{abipy} and abiflows\cite{abiflows}. 

The Monkhorst-Pack grids used to sample the Brillouin zone were distinguished considering the number of the points per reciprocal atom for both the electron wavefunctions (referred to as \textit{k}-points) and for the phonon wavefunctions (referred to as \textit{q}-points), respectively. These quantities, the number of points in the full Brillouin zone times the number of atoms, are labelled kpra and qpra, respectively.

For the statistical analysis, we identified, for each material, a reference grid (leading to converged phonon frequencies, see below). For each other sampling density the errors were defined as the absolute value of the difference between the phonon frequencies $\omega$ calculated with that grid and those obtained with the reference grid. In this context, when analyzing the \textit{k}-point convergence we considered the frequencies at the high symmetry points of the Brillouin zone\cite{Setyawan2010299}, while when varying the \textit{q}-points grid the convergence was checked for the interpolated values on very dense regular grids of \textit{q}-points (100000 qpra). In the latter case, \textit{q}-points belonging to the dense grids explicitly obtained from the DFPT calculation were excluded from the statistical analysis, as their error would be zero by definition.

For each specific grid we considered mainly the mean absolute error (MAE) and the mean absolute relative error (MARE) over all the frequencies. The choice of focusing on averaged errors stems from the fact that many quantities derived from phonon calculations that can be compared to experimental results are obtained integrating over the whole Brillouin zone (e.g. DOS and thermodynamic properties\cite{Lee1995}). In addition, in order to get a better insight on the presence of possible problematic regions, the maximum absolute error (MxAE) and the maximum absolute relative error (MxARE) were also considered. 

The reference configuration was selected independently for each material and for the \textit{k} and \textit{q} grids, by considering progressively increasing densities in the reciprocal space, until a convergence was reached based on the MAE and MARE.

\section{Electron Wavevector Grid Convergence}
\label{sec:kpt_conv}

\begin{figure}
\begin{center}
 \includegraphics[width=0.49\textwidth]{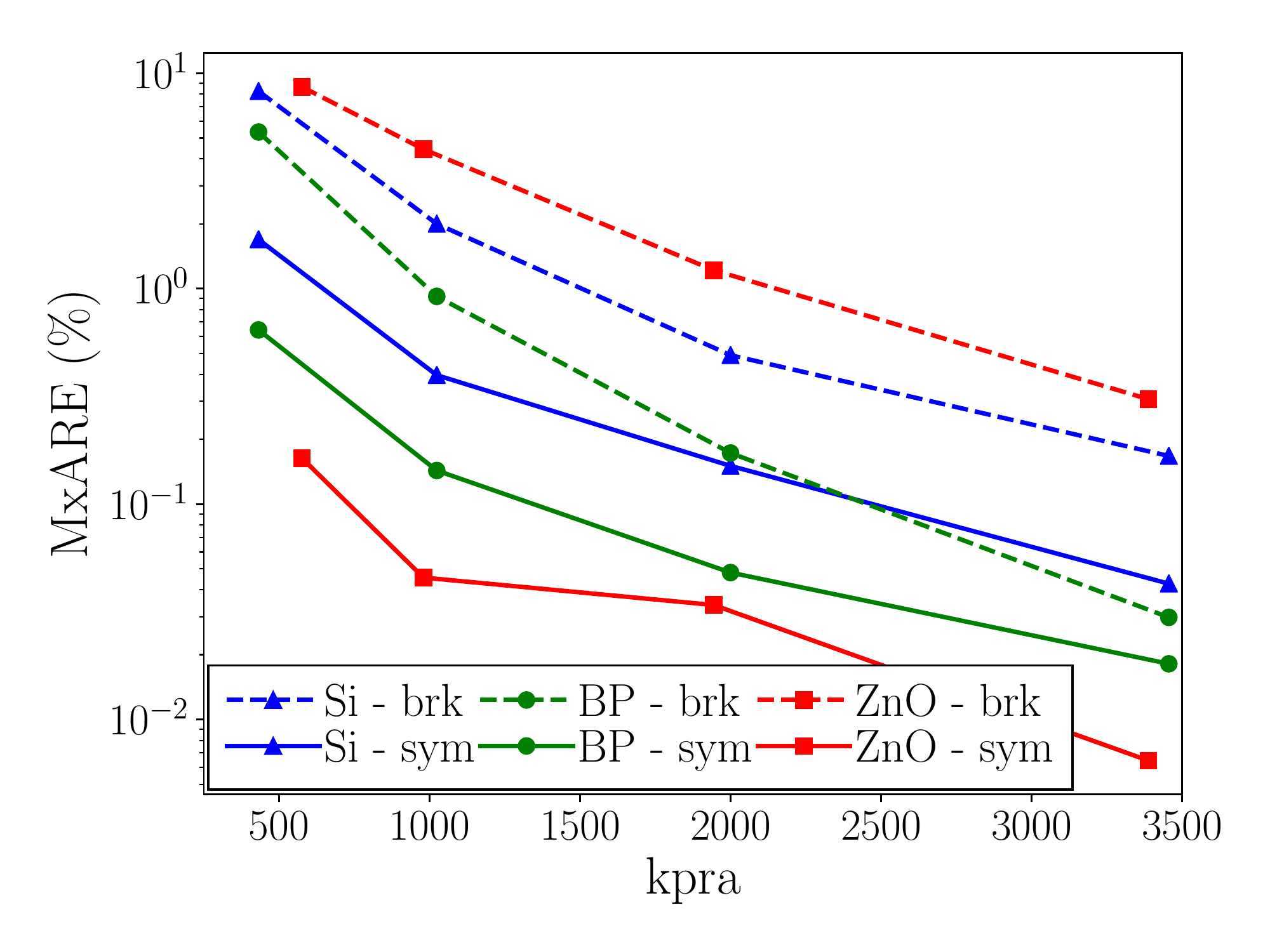}
 \caption{Evolution of the maximum relative error for different \textit{k}-point samplings for Si (blue triangle), BP (green circle) and ZnO (red square). The reference grids have approximately 5500 kpra. Solid lines represent the error for a $\Gamma$ centered grid, while dashed lines are for shifted grids, preserving and breaking the symmetries (sym and brk) of the crystals, respectively. The lines are a guide to the eye. \label{fig:csb}}
\end{center}
\end{figure}

Before any statistical analysis, we would like to highlight the importance of the choice of the k-point grid itself. For a subset of the materials considered (Si, BP and ZnO) the \textit{k}-point dependence of the phonon frequencies was investigated for a list of $\Gamma$ centered grids, thus satisfying the same symmetries of the crystal, and for the same grids with a shift that breaks the symmetry. The values of the MxARE with respect to a reference grid of approximately 5500 kpra are reported in Fig.~\ref{fig:csb}, showing that the symmetric set of grids can reduce the error by up to two orders of magnitude compared to the non-symmetric one. This demonstrates the importance of choosing a \textit{k}-point grid that respects the symmetries of the system to improve the rate of convergence. In the following, the \textit{k}-point grids were thus chosen with appropriate shifts satisfying this criterion. 

The relative and absolute errors for the \textit{k}-point grids were calculated as described in Section~\ref{sec:method} for each different grid available. Considering a threshold $\varepsilon$ for the maximum error tolerated and given a specific value $\overline{\text{kpra}}$ of the \textit{k}-points per reciprocal atom, we analyzed the fraction $F$ of materials for which the results are converged within the limits $\varepsilon$ and $\overline{\text{kpra}}$. More precisely, we define $F$ as:
\begin{equation}
\label{eq:fraction_conv}
F(\overline{\text{kpra}}, \varepsilon) = \frac{\sum\limits_{i=1}^{48}\left[\left|\text{Err}_i(\kappa_i)\right| < \varepsilon\right]}{48},
\end{equation} 
where $i$ labels the 48 materials, $\kappa_i$ is the largest grid available lower than $\overline{\text{kpra}}$
\begin{equation}
\kappa_i = \max_{\text{kpra}_i<\overline{\text{kpra}}}\text{kpra}_i,
\end{equation}
Err$_i(\kappa_i)$ represents either the MAE, MARE, MxAE or MxARE, while the Iverson bracket $[P]$ for a proposition $P$ is 1 if $P$ is true and 0 otherwise. 

\begin{figure}
\begin{center}
 \includegraphics[width=0.49\textwidth]{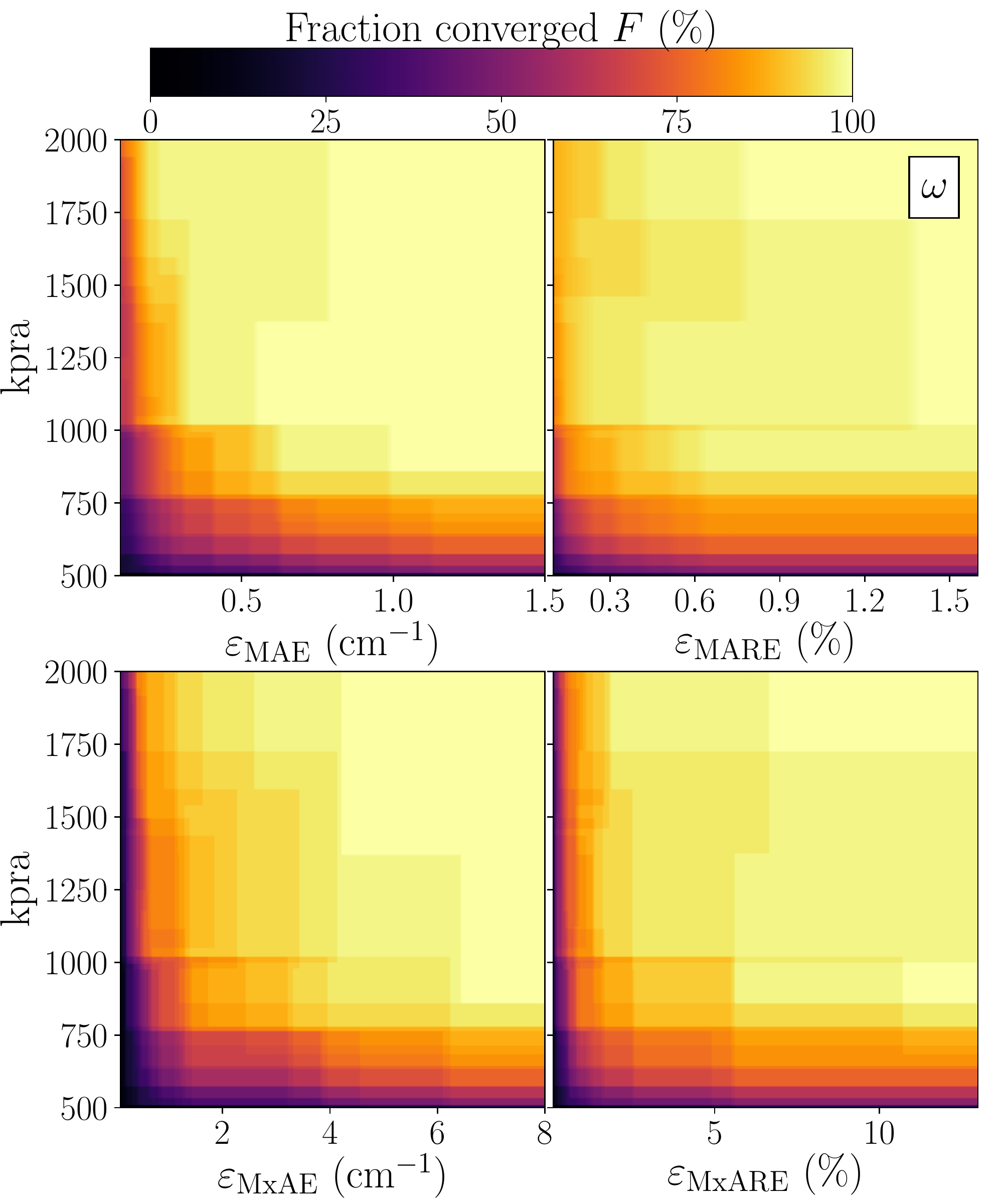}
 \caption{Fraction of materials converged $F$ (see Eq.~\eqref{eq:fraction_conv}) with respect to the phonon frequencies $\omega$ as a function of the \textit{k}-point density kpra and maximum error threshold allowed $\varepsilon$ for the different kind of errors MAE, MARE, MxAE, MxARE. \label{fig:err_kpra}}
\end{center}
\end{figure}

The values of $F$ as a function of kpra and $\varepsilon$ for the different types of errors are reported in Fig.~\ref{fig:err_kpra} and can be used to identify an appropriate value for the \textit{k}-point density. The mean errors, both relative and absolute, tend to converge rather fast and for grid denser than 1000 kpra all the materials can be considered converged with a MAE below $1$~cm$^{-1}$ and a MARE of $1.5\%$. However, this represents the largest mean error and, at this grid density 90\% of the materials are actually converged at a much lower error threshold (0.3~cm$^{-1}$ and 0.15\%).

It should be noted that $F$ is not strictly a monotonic function of kpra for a fixed value of $\varepsilon$, reflecting the oscillations in the error that show up in a few cases. These oscillations are eventually smoothed out for denser grids and, in the cases where this phenomenon is more relevant (AgCl and ScF$_3$), their origin can be traced back to a large error on a particular mode for a particular \textit{q}-point, suggesting that the impact on the overall band structure is quite limited. 

This can be seen also in the case of the maximum errors, where these large oscillations affects just a couple of cases, while almost all the materials show rather small values of MxRE and MxARE at 1000 kpra.

\begin{figure}
\begin{center}
 \includegraphics[width=0.3\textwidth]{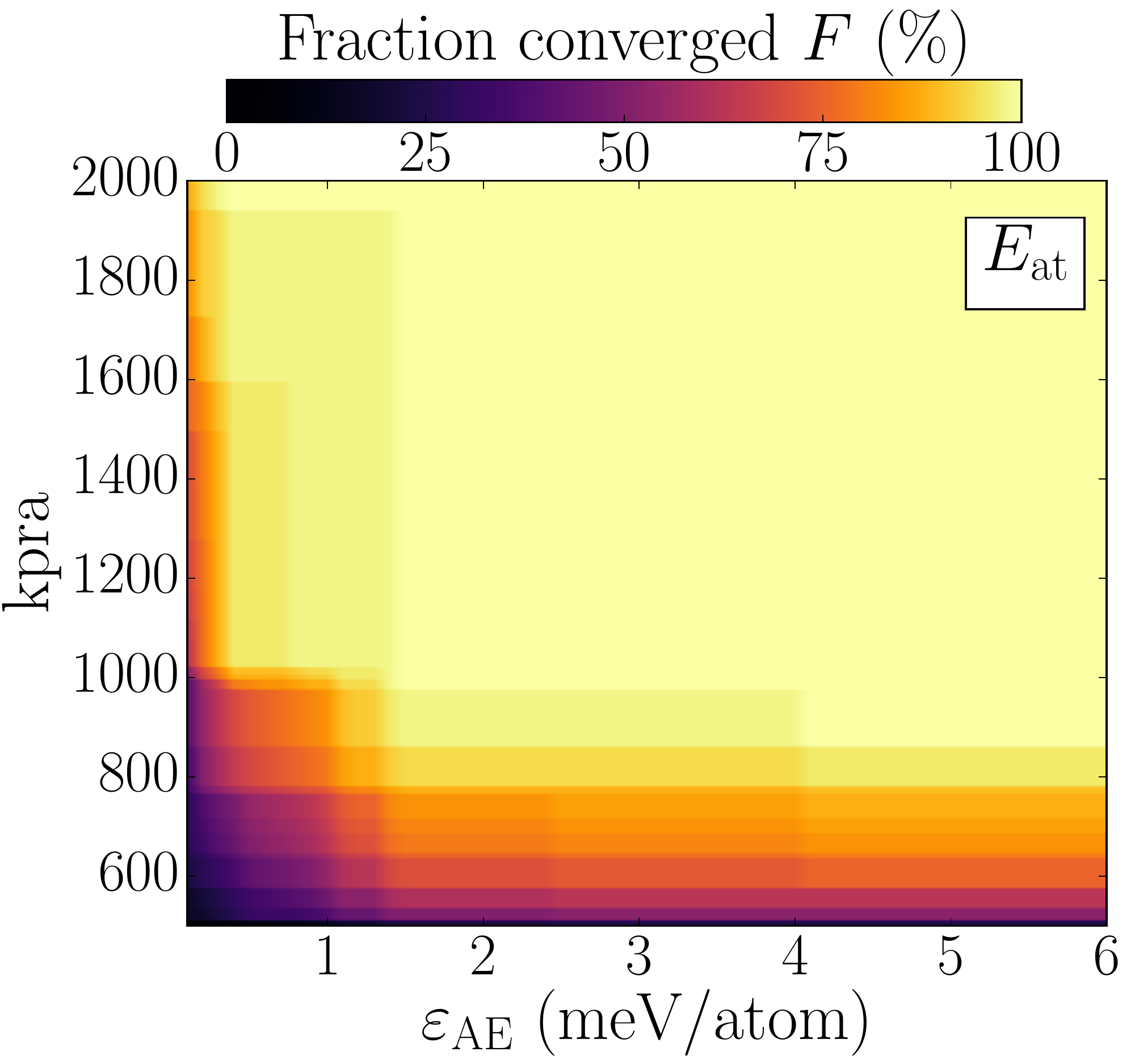}
 \caption{Fraction of materials converged $F$ (see Eq.~\eqref{eq:fraction_conv}) with respect to the energy per atom $E_{\text{at}}$ as a function of the \textit{k}-point density kpra and maximum error threshold allowed for the absolute error $\varepsilon_{\text{AE}}$. \label{fig:err_energy_kpra}}
\end{center}
\end{figure}

These results can be compared to the rate of convergence of the total energy per atom $E_{\text{at}}$. In Fig.~\ref{fig:err_energy_kpra} we report the values of $F$ for the absolute error (AE), showing how, with a kpra below 800, 95\% of the materials are converged within a threshold of 5~meV/atom. This demonstrates that, in general, a denser sampling of the Brillouin zone is required to obtained accurate perturbations, compared to the total energy.

In the case of polar materials, the $\Gamma$ point deserves further discussion. The error on the modes that have an LO-TO splitting is systematically larger compared to the error obtained neglecting the nonanalytical contribution. This is due to a slower convergence of the derivative with respect to the electric field. This should be usually taken into account when studying the phonon frequencies just at $\Gamma$, but it should also be mentioned that the maximum errors shown in Fig.~\ref{fig:err_kpra} are coming from modes at \textit{q}-points away from the origin for most of the materials. This means that when considering the whole Brillouin zone a denser \textit{k}-point grid is not needed for the electric field derivatives. 

\begin{figure}
\begin{center}
 \includegraphics[width=0.49\textwidth]{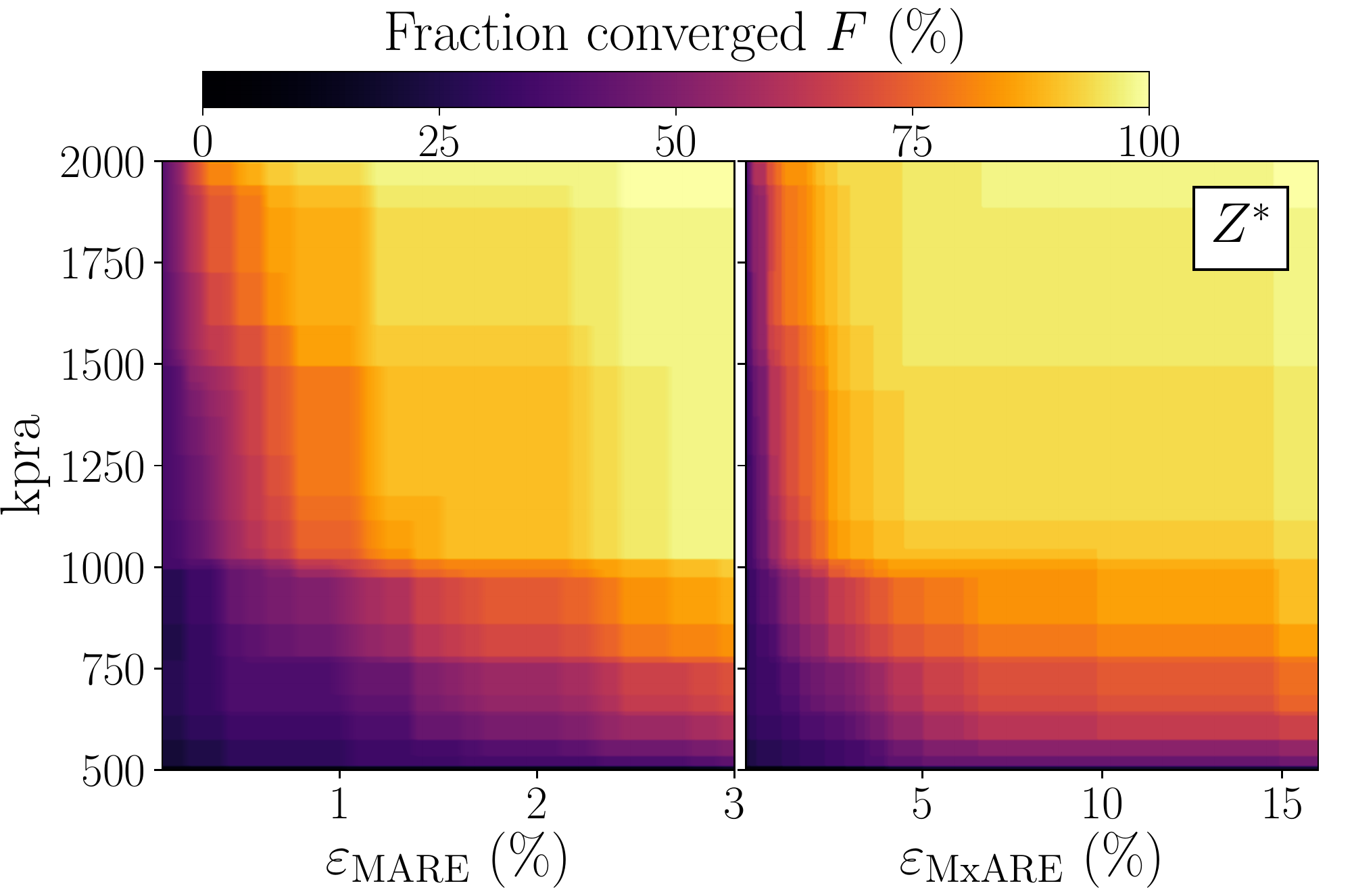}
 \caption{Fraction of materials converged $F$ (see Eq.~\eqref{eq:fraction_conv}) with respect to the BECs $Z^*$ as a function of the \textit{k}-point density kpra and maximum error threshold allowed $\varepsilon$ for the different kind of relative errors MARE and MxARE. \label{fig:err_becs_kpra}}
\end{center}
\end{figure}

\begin{figure}
\begin{center}
 \includegraphics[width=0.49\textwidth]{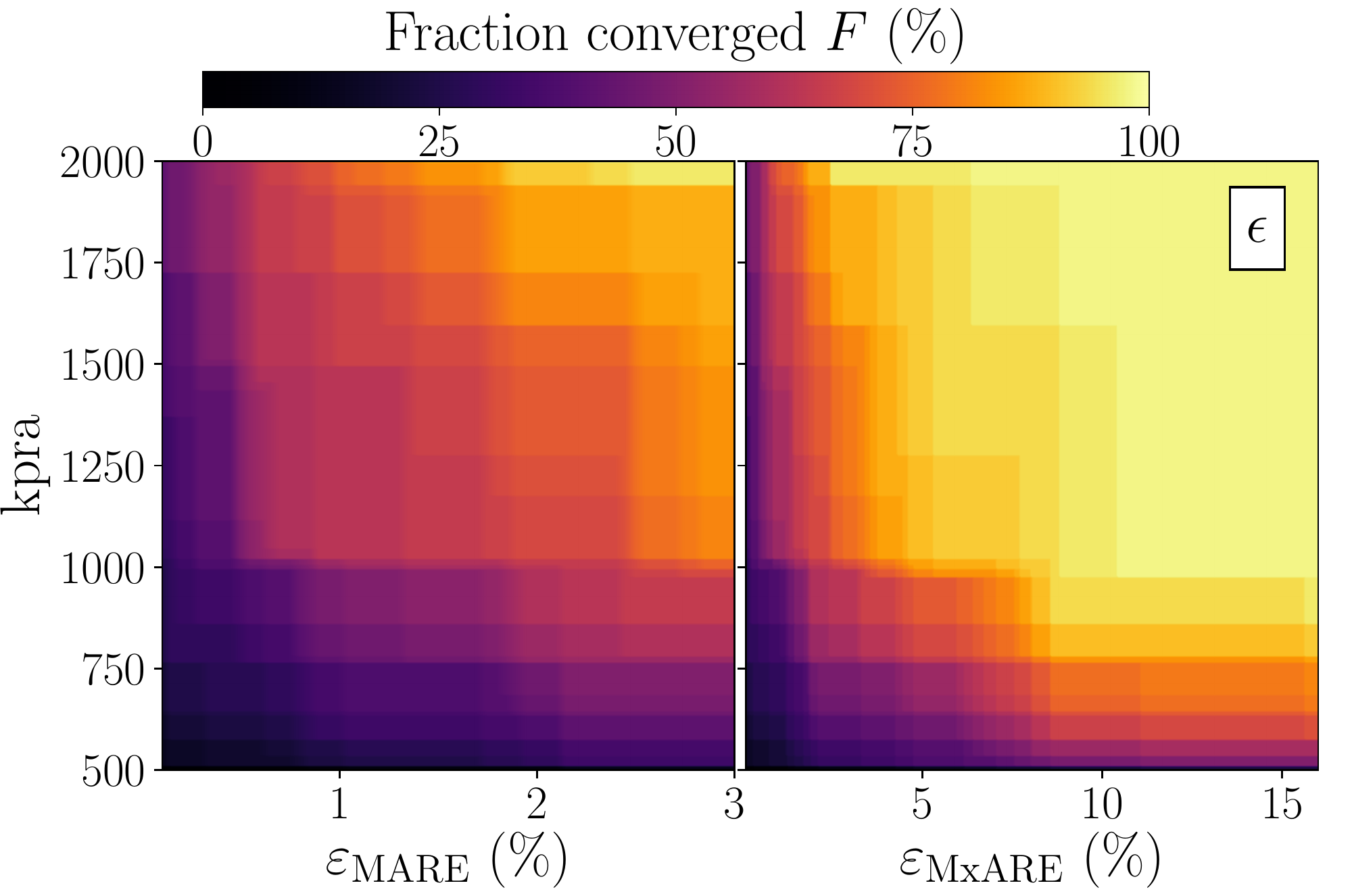}
 \caption{Fraction of materials converged $F$ (see Eq.~\eqref{eq:fraction_conv}) with respect to the dielectric tensor $\epsilon$ as a function of the \textit{k}-point density kpra and maximum error threshold allowed $\varepsilon$ for the different kind of relative errors MARE and MxARE. \label{fig:err_emacro_kpra}}
\end{center}
\end{figure}

As a further check, we explored the convergence of the BECs and dielectric constant $\epsilon$ for all the polar materials considered here. The fraction $F$ for the MARE and MxARE related to these two quantities are shown in Fig.~\ref{fig:err_becs_kpra} and Fig.~\ref{fig:err_emacro_kpra}, respectively. In general, the rate of convergence turned out to be slightly slower compared to the phonon frequencies shown in Fig.~\ref{fig:err_kpra}, but at 1000 kpra for these quantities roughly 85\% of the materials are converged with a tolerance of 5\% on the MxARE. This \textit{k}-point density is then suitable in general also for quantities purely related to the electric perturbations.

As a final point, we confirm that \textit{k}-points have little or no influence on the breaking of the ASR, when it is not imposed explicitly.

\section{Phonon Wavevector Grid Convergence}
\label{sec:qpt_conv}
Given the conclusions from Section~\ref{sec:kpt_conv}, we chose \textit{k}-point grids with densities larger than 1000 kpra for all the materials and the phonon frequencies were calculated using different sets of phonon wavevector (\textit{q}-points) samplings. The aim was to identify a value of qpra that would provide converged results for the phonon frequencies obtained through Fourier interpolation over the whole Brillouin zone.

\begin{figure}
\begin{center}
 \includegraphics[width=0.49\textwidth]{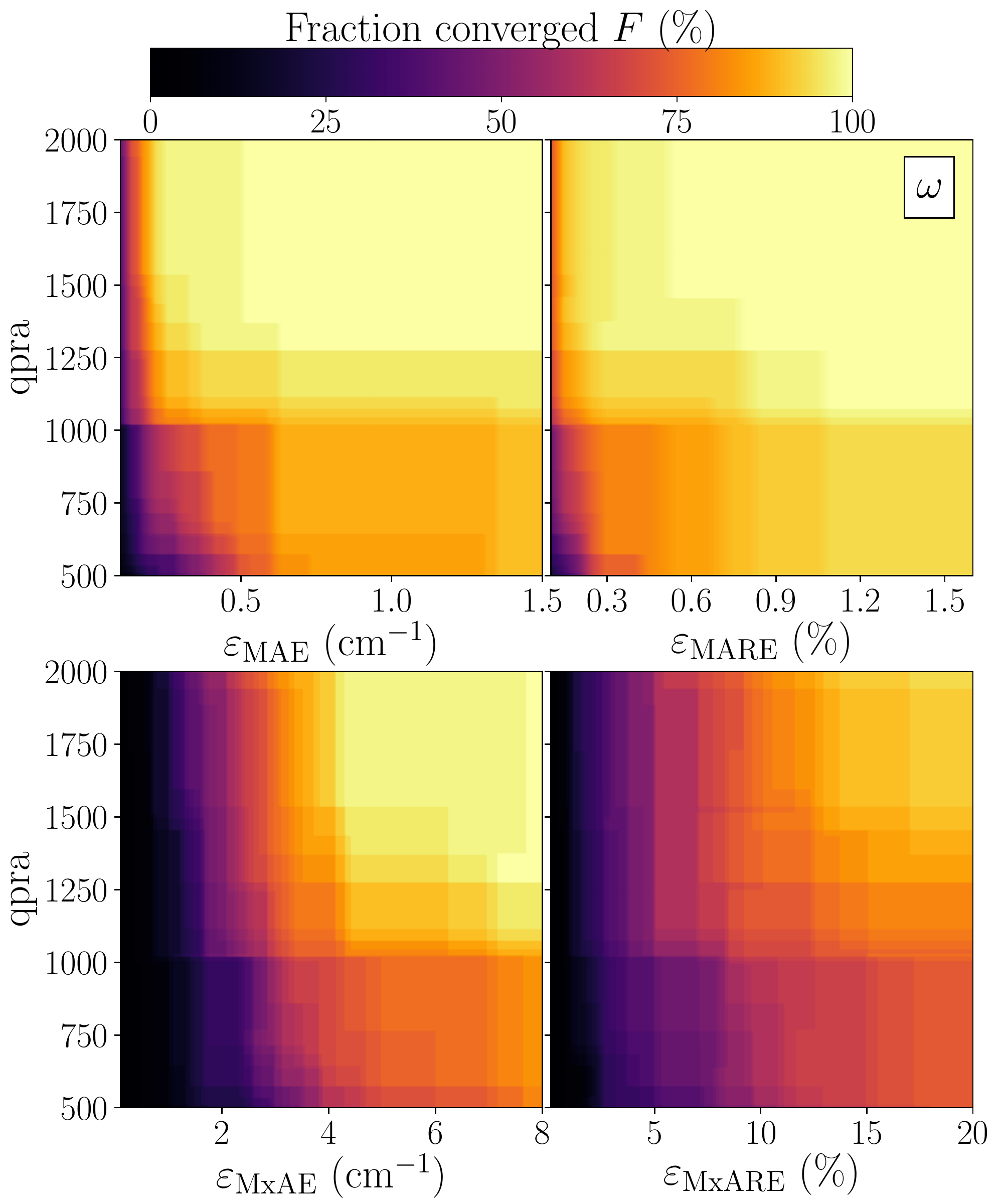}
 \caption{Fraction of materials converged $F$ (see Eq.~\eqref{eq:fraction_conv}) with respect to the phonon frequencies $\omega$ as a function of the \textit{q}-point density qpra and maximum error threshold allowed $\varepsilon$ for the different kind of errors MAE, MARE, MxAE, MxARE. \label{fig:err_qpra}}
\end{center}
\end{figure}

The values of the fraction of converged materials $F$, as defined in Eq.~\eqref{eq:fraction_conv}, are shown in Fig.~\ref{fig:err_qpra} for the MAE, MARE, MxAE and MxARE as a function of qpra and $\varepsilon$. Comparing these results with the case of the \textit{k}-point sampling, it can be seen immediately that the rate of convergence with respect to the \textit{q}-point density is slower and in particular that there is a strikingly different behaviour of the maximum errors.

The presence of large maximum errors can be partially understood on the basis of two considerations: first, the fact that values are obtained through an interpolation could introduce errors related to the interpolation scheme. Second, evaluating the errors on a very large set of points increases the probability of coming across a small region of the Brillouin zone that is poorly described, possibly leading to a large maximum error. Under these assumptions it is reasonable to expect that the average error will not be increased substantially.

However, while these factors definitely play a role, analyzing each material individually, we also identified a series of problems giving a strong contribution to the maximum error. 

One of these problems, present in different materials, shows up when the \textit{q}-point grid becomes denser than the \textit{k}-point grid. The frequencies tend to deviate from the linear trend that is expected for the acoustic modes close to $\Gamma$. The problem becomes more and more evident as the \textit{q} grid gets denser. Due to the interpolation, these deviations tend to propagate also further away from the close proximity of $\Gamma$ (see Fig~S1 of the SM for an example). This can lead to a large maximum error, as we have defined it, due to an incorrect value of the reference frequencies. In particular, since this involves mainly small frequencies for few modes and few \textit{q}-points, the MxARE tends to be very large, while the average errors still keeps improving with increasing qpra. Despite this, we explicitly verified in the cases of K$_2$O, FeS$_2$, RbI and RbYO$_2$, that changing the \textit{k}-point grid to match exactly the values of the \textit{q}-point grid eliminates completely these deviations. These calculations show that the frequencies (even close to $\Gamma$) were already well converged even with a lower qpra (see Fig.~S2 of the SM).

Although we were not able to identify the origin of the problem univocally there are various possible sources of error for $\mathbf{q}\rightarrow0$. First the wavelength of the phonon perturbations is longer than the electronic one. There is also the need for additional calculations to handle incommensurate grids. Moreover, we have observed that adding an appropriate cutoff in real space on the IFCs corrects the behavior close to $\Gamma$ (see Fig.~S3 of the SM), suggesting that some numerical inaccuracies at long wavelength should be involved.

We point out that the value of such a cutoff on the IFCs should be tuned to get satisfactory results of the phonon frequencies. So, while this may be a viable option to improve the results in standard phonon calculations, its use in an high-throughput process is limited by the need of an automated determination of a suitable value of the cutoff, that may be not trivial.

A second problem appears in AgCl. On top of some inaccuracies as those discussed above, this material has its relaxed PBE lattice parameters close to a phase transition\cite{Li2006}. As a consequence some acoustic modes are softened along the $\Gamma$-K direction and at L, with the onset of small dips in the phonon band structure (see Fig.~S4 of the SM). These irregularities may be not properly described with a Fourier interpolation on a regular grid of \textit{q}-points, so large MxRE and MxARE arise in those zones due to the different interpolations obtained with different \textit{q}-point grids.

Aside from these problems, we also highlight the fact that deviations from the linear trend for acoustic modes at $\Gamma$ could signal a lack of convergence of the energy cutoff or the \textit{k}-point sampling, and that can be used as an indicator to identify unconverged calculations during an high-throughput screening.

Based on these considerations about the maximum errors and on the results of Fig.~\ref{fig:err_qpra}, we conclude that nonetheless a larger  qpra is needed in general with respect to the kpra, with an optimal value set at around 1500 points per reciprocal atom. With this value all the materials could be considered converged within a MAE of 0.5 cm$^{-1}$ and a MARE below 0.6\%. Such a qpra value also leads to satisfactory maximum errors, with the exception of the cases discussed above.  

Due to the problems highlighted above, it would also be convenient to raise the value of kpra as well, in order to always work with \textit{q}-point and \textit{k}-point grids of the same size.

The conclusion of the present study also applies to the computation of phonon frequencies using the supercell method\cite{Kresse1995,Kunc1982}, Fourier extrapolated using interatomic force constants. Indeed, the present DFPT approach and the supercell method do not differ at the level of their interpolation scheme. This means that the above-mentioned MAE and MARE targets or maximum errors need supercells of 1500 atoms to be attained. The current practice with supercell calculations, 
however, usually rely on supercells with less than 200 atoms. Hence, the accuracy of such studies is quite limited.
This illustrates the power of the DFPT approach, that uses only the primitive cell, even for incommensurate q-point wavevectors.

\section{Validation}
\label{sec:validation}

To validate our set of results and highlight the effects of the problems discussed in the previous sections further checks have been performed. One aimed at the analysis of the pathological region in the proximity of $\Gamma$ and the other aimed at testing the accuracy in comparison with experimental data.

\subsection{Sound velocity}
\label{subsec:sound}
The first test is based on the calculation of the sound velocity of the materials, as obtained from the slope of the acoustic mode close to the $\Gamma$ point. Since not many experimental data are available, here we focus on establishing the precision of our results, rather then the accuracy. We then calculated the same sound velocities from the DFPT calculations of the elastic constants and the help of the Christoffel tensor. These quantities should match for converged calculations and allow to verify the trends close to the origin of the Brillouin zone. In fact, apart from the problems mentioned in Section~\ref{sec:qpt_conv}, the linear behavior of the acoustic modes can eventually break down for small \textit{q}, altering the value of the slope extracted from the interpolation.

\begin{figure}
\begin{center}
 \includegraphics[width=0.49\textwidth]{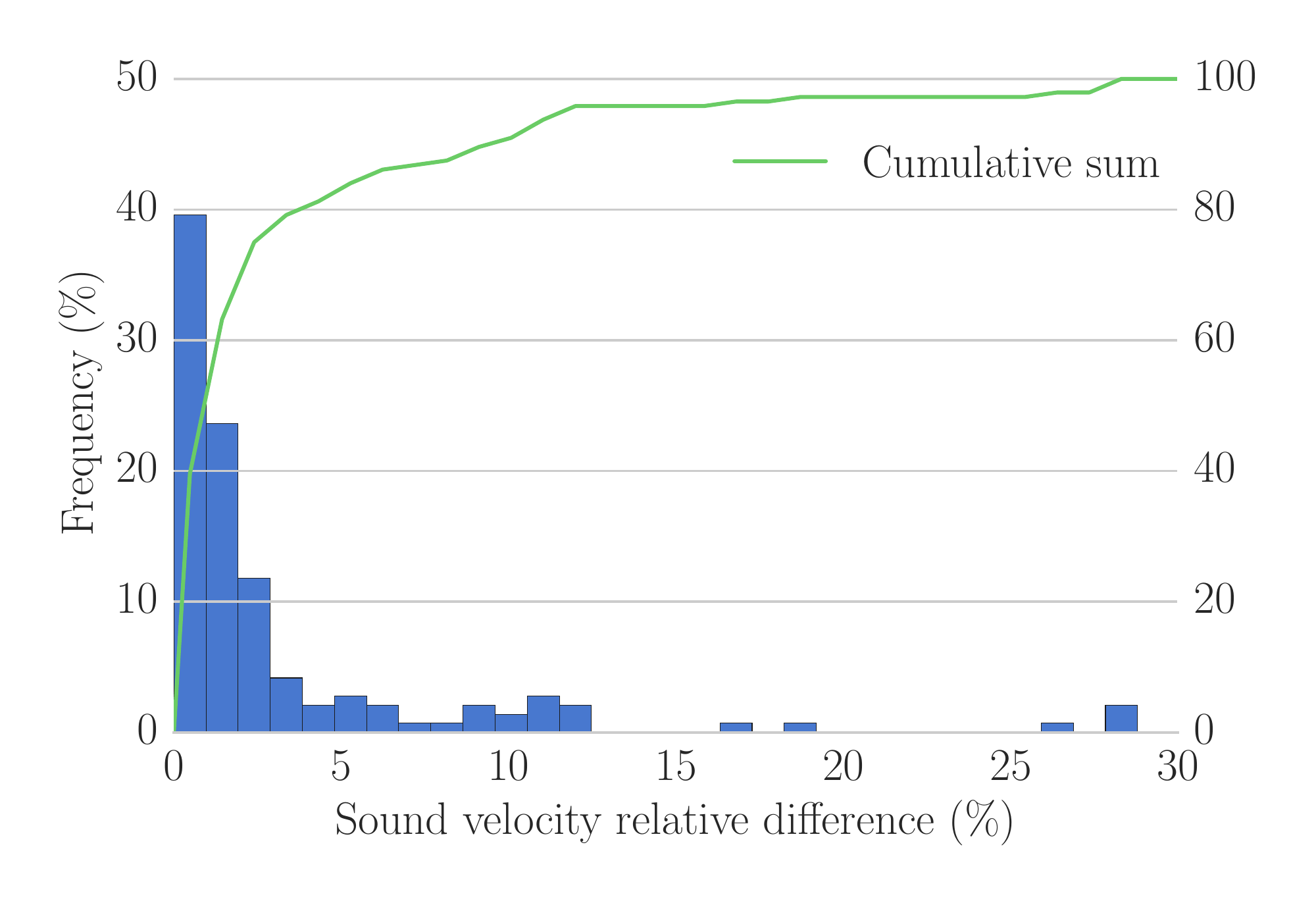}
 \caption{Distribution of the relative difference between the sound velocities obtained from the slope of the acoustic modes and from the elastic tensor. Each values considered is for one of the three acoustic modes, averaged over the three cartesian direction. \label{fig:sound_vel}}
\end{center}
\end{figure}

More in detail, elastic constants were calculated with the same approximations used to run the phonon simulations in Section~\ref{sec:qpt_conv}, while to extract the sound velocity from phonon frequencies \textit{q}-point grids of approximately 1500 qpra were used. We compared the values of the speed of sound for the three acoustic modes averaged over the three cartesian directions. The distribution of the relative difference between the two different types of calculations is shown in Fig.~\ref{fig:sound_vel}.

The agreement is generally good, with more than 80\% of the evaluated sound velocities having a difference lower than 5\% between the two methods and more then 90\% below 10\% difference. 

The outliers with large differences can be easily understood. The first is ZnSe, for which the cutoff energy is not large enough for the calculation of the elastic tensor, while phonon bands are converged. As the cutoff is increased the agreement is in line with other materials. The second is AgCl, that was discussed in the previous section and that shows problems close to $\Gamma$ coming from the \textit{q}-point sampling and from the presence of soft modes. This is likely to lead to inaccuracies in the interpolated values of the sound velocity. 

The other smaller differences can be rationalized in terms of the convergence with respect to the k-points, q-points and energy cutoff  as well as the small instabilities close to $\Gamma$. 
We can thus conclude that the precision achieved with the suggested densities is satisfactory even for the description of the central region of the Brillouin zone, in particular having high-throughput calculations in mind.

\subsection{Thermodynamic properties}
\label{subsec:thermo}
The second part of the validation concerns the precision achieved in comparison with experimental data. In particular we used as a target the values of the entropy $S$ at room temperature (i.e. around 300 K, matching the experimental data available). This can be obtained from the normalized phonon density of states $g(\omega)$ \cite{Lee1995}:
\begin{eqnarray}
\label{eq:entropy}
S(T) = 3nNk_B\int_{0}^{\omega_L}\left(\frac{\hbar\omega}{2k_BT}\text{coth}\left(\frac{\hbar\omega}{2k_BT}\right) \right. \nonumber \\* 
 \left. - \text{ln}\left(2\text{sinh}\left(\frac{\hbar\omega}{2k_BT}\right)\right)\right)g(\omega)d\omega
\end{eqnarray}
where the integral is carried out over all the phonon frequencies, $k_B$ is the Boltzmann constant, $n$ is the number of atoms per unit cell, $N$ is the number of unit cells,  and $T$ is the temperature.

Entropy provides a good example of quantity integrated over the whole Brillouin zone, that should have errors comparable to those observed for MAE and MARE discussed in Section~\ref{sec:qpt_conv}. The rate of convergence with respect to \textit{q}-points is indeed quite fast and for a density of 1500 qpra the maximum relative error observed for the entropy at room temperature with respect to the reference \textit{q}-point grid is below 0.7\% for all the compounds. 

\begin{figure}
\begin{center}
 \includegraphics[width=0.49\textwidth]{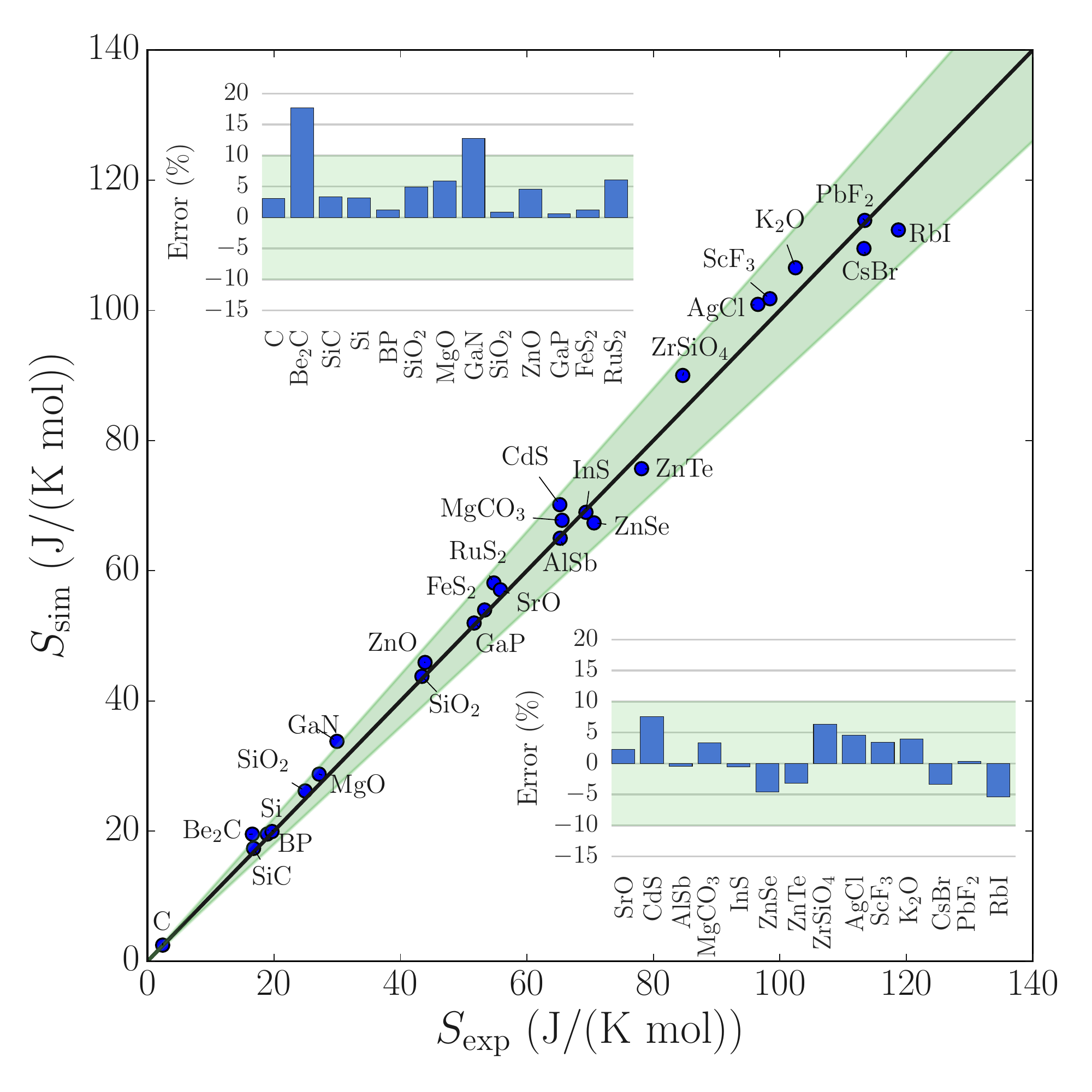}
 \caption{Experimental vs theoretical data for entropy $S$ at room temperature. The green area delimits a region with relative error below 10\%. The insets represent the relative error with respect to the experimental value.\label{fig:entropy_exp}}
\end{center}
\end{figure}

These values of the entropy were compared with the experimental data for the 27 compounds for which these are available \cite{Barin2008,hemingway1977,Lord2004,JANAF,LiHuang2004,HOLM19672289}. This comparison is shown in Fig.~\ref{fig:entropy_exp}, where the relative errors are reported as well. 

The agreement is generally good, considering that other sources of entropy can enter in the experimental values, with the exception of Be$_2$C and GaN. The MARE between the theoretical and experimental values is 4.2\%, in line with data present in literature\cite{He2014,MA20153762,gurel2010,ZHANG2010976}, while the MAE is 2.23~J/(K~mol) ($9.6\times10^{-3}$~meV/(K~atom)). Interestingly, at 300~K this is much smaller than the estimated error of other energy contributions for the estimation of phase stability\cite{Hautier2011b}. It should be noted that at room temperature the anharmonic contribution to the entropy, that were not taken into account in our calculations, could partially explain the discrepancies and the systematic overestimation of the theoretical results.

These results allows us to confirm the reliability of our calculations, with a good estimate of the precision in terms of integrated quantities.

\section{Conclusion}

We have studied the convergence of the phonon frequencies obtained in the framework of DFPT as a function of the sampling of the Brillouin zone, in order to provide a reliable high-throughput framework for the calculation of phonon band structures. 

We have highlighted the common problems emerging from the choices of the sampling densities. In particular, for the \textit{k}-points, a higher density is needed for obtaining well-converged values of the LO-TO splitting compared to the simple phononic perturbations at $\Gamma$. A slightly higher density is also generally needed for \textit{q}-points other than the origin.

Concerning the frequencies obtained from the interpolation of a regular \textit{q} grid, the region close to $\Gamma$ is the most problematic due to numerical inaccuracies (insufficiently converged parameters, long range oscillations), proximity of the system to a phase transition and numerical inaccuracies in the Fourier interpolation. 

It has also been observed that imposing a cutoff on the IFCs in real space may help in solving problems related to numerical inaccuracies originating from different sources. However, an appropriate value of the cutoff should be chosen individually for each material and the results should be thoroughly checked. 

In general, a smaller density of 1000 points per reciprocal atoms seems to be necessary for the \text{k}-points compared to the qpra. However, having observed some pathological behavior close to the origin of the Brillouin zone when using \textit{q} grids denser than \textit{k} grids, we suggest that the best approach would be to use \textit{k} grids with an appropriate single shift that preserves the symmetries of the crystal along with equivalent \textit{q} grids, both with a density of approximately 1500 points per reciprocal atom. This should provide well converged results for the majority of semiconducting materials.

While these hints were extracted to tune the calculations in an high-throughput regime, they can also be used as indications for standard phonon calculations.

\section*{Acknowledgement}
We gratefully acknowledge discussions with B.~Van~Troeye and M.~Giantomassi. G.P., X.G. and G.-M.R. acknowledge support from the Communaut\'{e} fran\c{c}aise de Belgique through the BATTAB project (ARC 14/19-057). G.-M.R. is also grateful to the F.R.S.-FNRS for the financial support. Computational resources have been provided by the supercomputing facilities of the Universit\'{e} catholique de Louvain (CISM/UCL) and the Consortium des Equipements de Calcul Intensif en F\'{e}d\'{e}ration Wallonie Bruxelles (CECI) funded by the Fonds de la Recherche Scientifique de Belgique (F.R.S.-FNRS).

\bibliographystyle{elsarticle-num}
\bibliography{bibliography}
\end{document}


\title{Convergence and pitfalls of density functional perturbation theory phonons calculations from a high-throughput perspective}

\author{Guido Petretto}
\affiliation{Universit\'{e} Catholique de Louvain, Institute of Condensed Matter and Nanosciences 
(IMCN)}

\author{Geoffroy Hautier}
\affiliation{Universit\'{e} Catholique de Louvain, Institute of Condensed Matter and Nanosciences 
(IMCN)}

\author{Xavier Gonze}
\affiliation{Universit\'{e} Catholique de Louvain, Institute of Condensed Matter and Nanosciences 
(IMCN)}

\author{Gian-Marco Rignanese}
\affiliation{Universit\'{e} Catholique de Louvain, Institute of Condensed Matter and Nanosciences 
(IMCN)}

\maketitle

\section{Materials}
The details of the materials considered for the current study are listed in Table \ref{tab:materials_data}. The mp-id is the Materials Project identifier, and the band gaps reported have been obtained with same approximations as those described in the main text.

\begin{table*}
  \centering
  \caption{List of the materials with their respective properties. The mp-id key is the Materials Project identifier.}
  \label{tab:materials_data}
  \begin{tabularx}{0.9\textwidth}{|Y|Y|Y|Y|Y|Y|Y|}
    \hline
    mp-id & Formula & N sites & Crystal system & Space group & ecut (Ha) & Band gap (eV) \\
\hline
\hline
mp-66 & C & 2 & cubic & 227 & 35 & 4.17 \\
\hline 
mp-149 & Si & 2 & cubic & 227 & 20 & 0.61 \\
\hline 
mp-1265 & MgO & 2 & cubic & 225 & 48 & 4.40 \\
\hline 
mp-1479 & BP & 2 & cubic & 216 & 30 & 1.26 \\
\hline 
mp-2176 & ZnTe & 2 & cubic & 216 & 35 & 1.07 \\
\hline 
mp-2472 & SrO & 2 & cubic & 225 & 48 & 3.33 \\
\hline 
mp-2624 & AlSb & 2 & cubic & 216 & 26 & 1.22 \\
\hline 
mp-2667 & CsAu & 2 & cubic & 221 & 38 & 1.09 \\
\hline 
mp-8062 & SiC & 2 & cubic & 216 & 35 & 1.38 \\
\hline 
mp-22903 & RbI & 2 & cubic & 225 & 33 & 3.77 \\
\hline 
mp-22922 & AgCl & 2 & cubic & 225 & 39 & 0.92 \\
\hline 
mp-571222 & CsBr & 2 & cubic & 225 & 38 & 4.27 \\
\hline 
mp-315 & PbF2 & 3 & cubic & 225 & 53 & 4.44 \\
\hline 
mp-971 & K2O & 3 & cubic & 225 & 48 & 1.73 \\
\hline 
mp-1569 & Be2C & 3 & cubic & 225 & 35 & 1.19 \\
\hline 
mp-10182 & LiZnP & 3 & cubic & 216 & 43 & 1.34 \\
\hline 
mp-12558 & LiMgAs & 3 & cubic & 216 & 43 & 1.36 \\
\hline 
mp-380 & ZnSe & 4 & hexagonal & 186 & 37 & 1.18 \\
\hline 
mp-672 & CdS & 4 & hexagonal & 186 & 39 & 1.12 \\
\hline 
mp-804 & GaN & 4 & hexagonal & 186 & 48 & 1.76 \\
\hline 
mp-2074 & Li3Sb & 4 & cubic & 225 & 43 & 0.69 \\
\hline 
mp-2133 & ZnO & 4 & hexagonal & 186 & 48 & 0.81 \\
\hline 
mp-5077 & NaLi2Sb & 4 & cubic & 225 & 43 & 0.68 \\
\hline 
mp-8882 & GaP & 4 & hexagonal & 186 & 48 & 1.29 \\
\hline 
mp-10694 & ScF3 & 4 & cubic & 221 & 53 & 6.19 \\
\hline 
mp-14437 & RbYO2 & 4 & trigonal & 166 & 48 & 3.66 \\
\hline 
mp-8278 & Ba(MgP)2 & 5 & trigonal & 164 & 29 & 1.16 \\
\hline 
mp-8280 & Ba(MgAs)2 & 5 & trigonal & 164 & 30 & 1.09 \\
\hline 
mp-9564 & Ca(MgAs)2 & 5 & trigonal & 164 & 30 & 1.26 \\
\hline 
mp-9570 & Ca(CdP)2 & 5 & trigonal & 164 & 39 & 0.83 \\
\hline 
mp-1522 & FeS2 & 6 & orthorhombic & 58 & 67 & 0.82 \\
\hline 
mp-6947 & SiO2 & 6 & tetragonal & 136 & 48 & 5.11 \\
\hline 
mp-13276 & SrLiP & 6 & hexagonal & 194 & 43 & 1.37 \\
\hline 
mp-2961 & MgSiP2 & 8 & tetragonal & 122 & 29 & 1.37 \\
\hline 
mp-3762 & VCu3S4 & 8 & cubic & 215 & 58 & 1.01 \\
\hline 
mp-19795 & InS & 8 & orthorhombic & 58 & 38 & 1.48 \\
\hline 
mp-36508 & SnHgF6 & 8 & trigonal & 148 & 53 & 2.43 \\
\hline 
mp-864954 & MgMoN2 & 8 & hexagonal & 194 & 34 & 0.72 \\
\hline 
mp-6930 & SiO2 & 9 & trigonal & 154 & 48 & 5.71 \\
\hline 
mp-14983 & Si4P4Ru & 9 & triclinic & 1 & 38 & 1.43 \\
\hline 
mp-5348 & MgCO3 & 10 & trigonal & 167 & 48 & 5.02 \\
\hline 
mp-989407 & YWN3 & 10 & orthorhombic & 26 & 32 & 1.61 \\
\hline 
mp-2030 & RuS2 & 12 & cubic & 205 & 38 & 0.57 \\
\hline 
mp-4820 & ZrSiO4 & 12 & tetragonal & 141 & 48 & 4.60 \\
\hline 
mp-10155 & P2Ir & 12 & monoclinic & 14 & 23 & 0.67 \\
\hline 
mp-2979 & ZnGeN2 & 16 & orthorhombic & 33 & 35 & 1.70 \\
\hline 
mp-570844 & Ga3Os & 16 & tetragonal & 136 & 48 & 0.68 \\
\hline 
mp-567841 & Be3P2 & 40 & cubic & 206 & 23 & 0.89 \\
\hline 
  \end{tabularx}
\end{table*}

\section{q-points convergence}

\begin{figure*}
\begin{center}
 \includegraphics[width=0.74\textwidth]{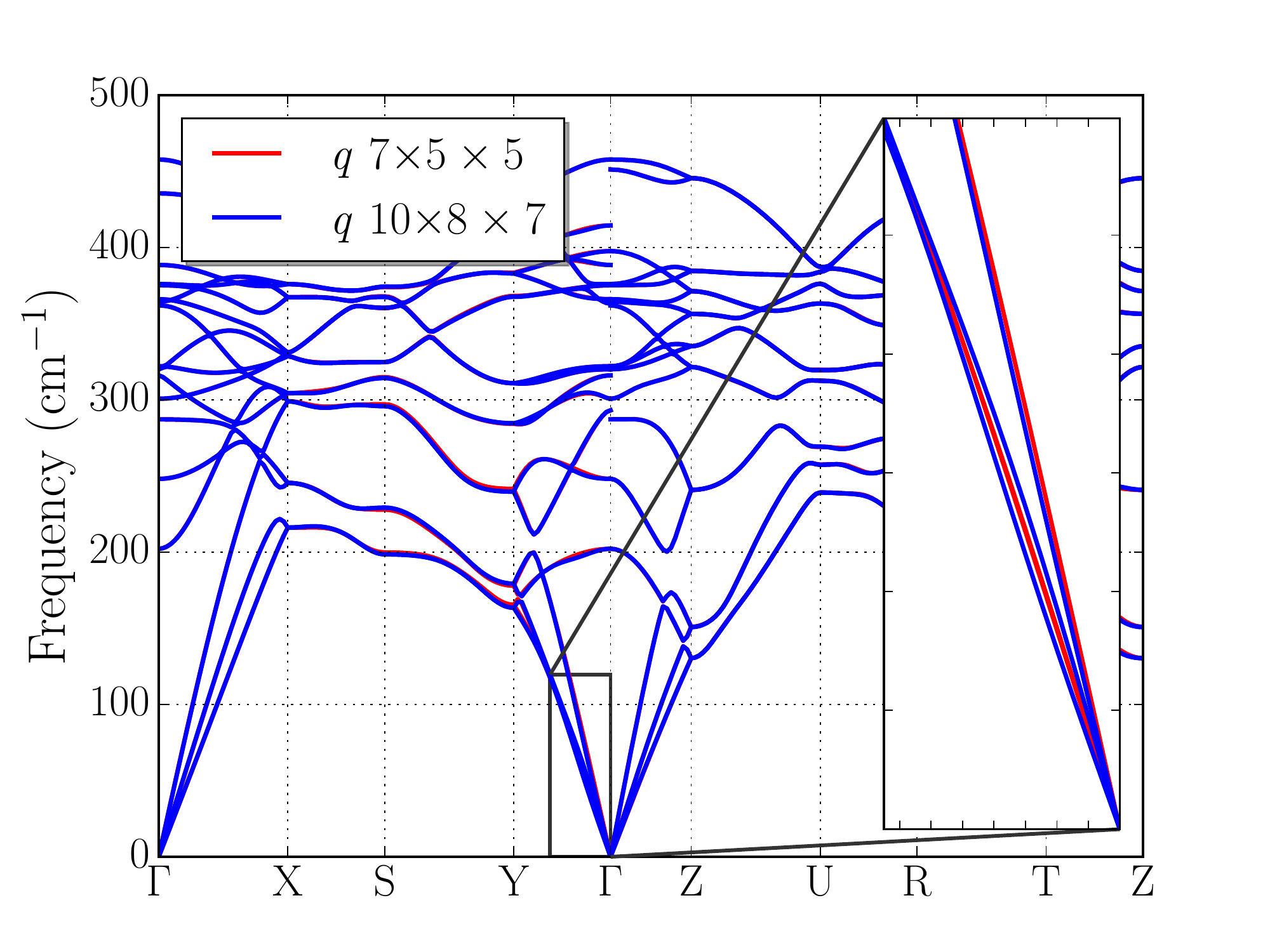}
 \caption{FeS2 phonon band structure with a $7\times5\times5$ \textit{k}-point grid and with $7\times5\times5$ and $10\times8\times7$ \textit{q}-point grid for red and blue lines, respectively. \label{fig_sm:FeS2_bs_oscillations}}
\end{center}
\end{figure*}

\begin{figure*}
\begin{center}
 \includegraphics[width=0.74\textwidth]{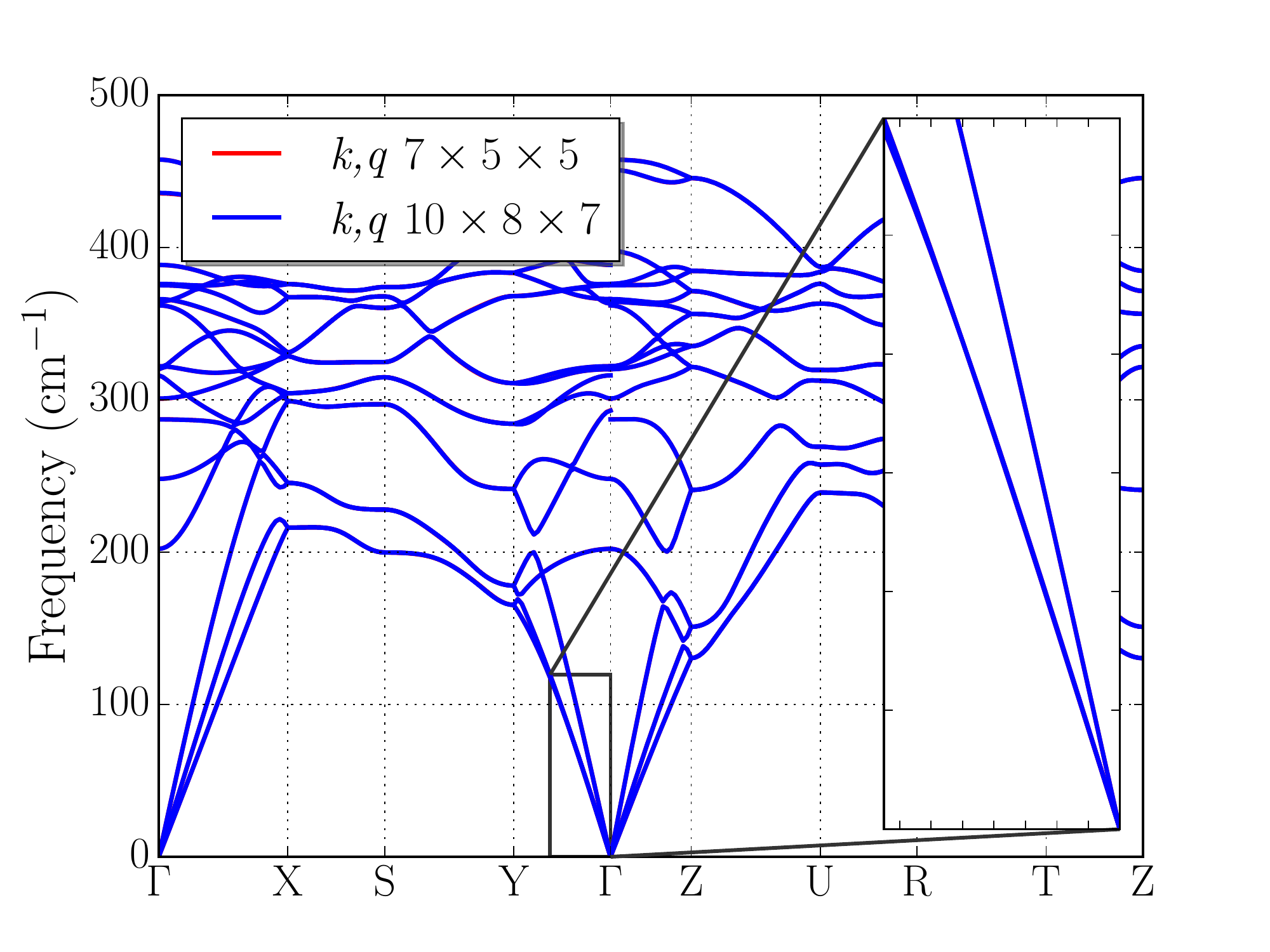}
 \caption{FeS2 phonon band structure with equal \textit{k}-point and \textit{q}-point grids $7\times5\times5$ and $10\times8\times7$  grid for red and blue lines, respectively. The lines are almost completely overlapping. \label{fig_sm:FeS2_bs_equiv}}
\end{center}
\end{figure*}

The convergence of the phonon band structures with respect to \textit{q} grid density has highlighted how the usage of a \textit{q}-point grid denser than the \textit{k}-point grid may lead to numerical instabilities close to the $\Gamma$ point. This is demonstrated in Fig. \ref{fig_sm:FeS2_bs_oscillations} and Fig. \ref{fig_sm:FeS2_bs_equiv}, where the phonon band structure of FeS$_2$ (mp-1522) with different sampling are shown. 

In Fig. \ref{fig_sm:FeS2_bs_oscillations} a common \textit{k}-point grid $7\times5\times5$ is used and the results obtained from the interpolation of $7\times5\times5$ and $10\times8\times7$ \textit{q}-point grids are compared, highlighting partial lost of a linear trend for the acoustic modes. At variance, when the \textit{k}-point grid is increased to match the \textit{q}-point one, the interpolated phonon frequencies are in excellent agreement (Fig. \ref{fig_sm:FeS2_bs_equiv}).

\begin{figure*}
\begin{center}
 \includegraphics[width=0.74\textwidth]{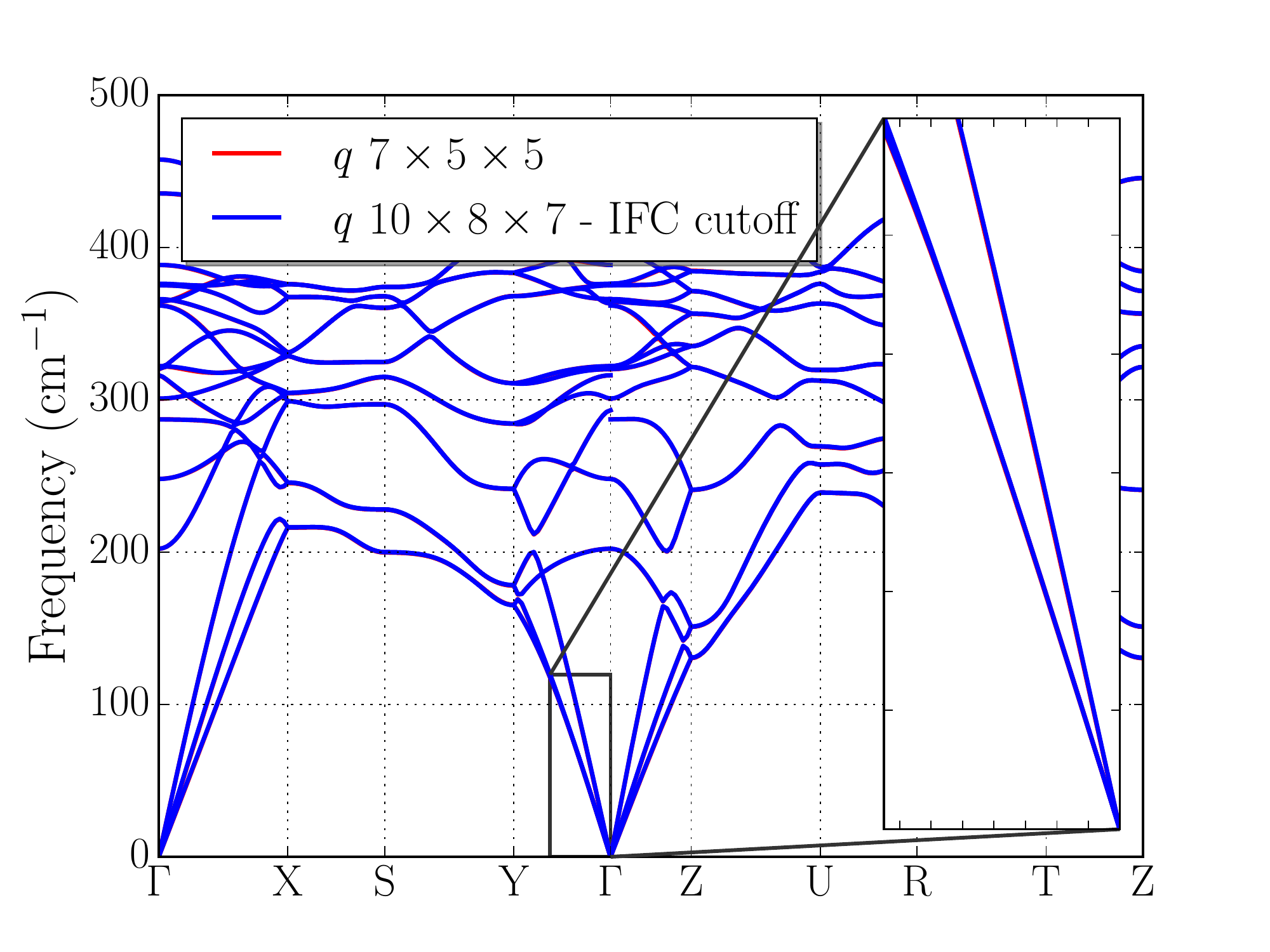}
 \caption{FeS2 phonon band structure with a $7\times5\times5$ \textit{k}-point grid and with $7\times5\times5$ and $10\times8\times7$ \textit{q}-point grid for red and blue lines, respectively. A cutoff of 25 Bohr on the IFCs has been imposed in the interpolation of the latter grid. The lines are almost completely overlapping.\label{fig_sm:FeS2_bs_ifc_cutoff}}
\end{center}
\end{figure*}

The same outcome can be obtained in this case by adding a cutoff of 25 Bohr on the interatomic force constants (IFCs) during the Fourier interpolation, even when considering the fixed $7\times5\times5$ \textit{k}-point grid, as shown in Fig. \ref{fig_sm:FeS2_bs_ifc_cutoff}.

The same behavior has been observed for K$_2$O, RbI and RbYO$_2$.

\begin{figure*}
\begin{center}
 \includegraphics[width=0.74\textwidth]{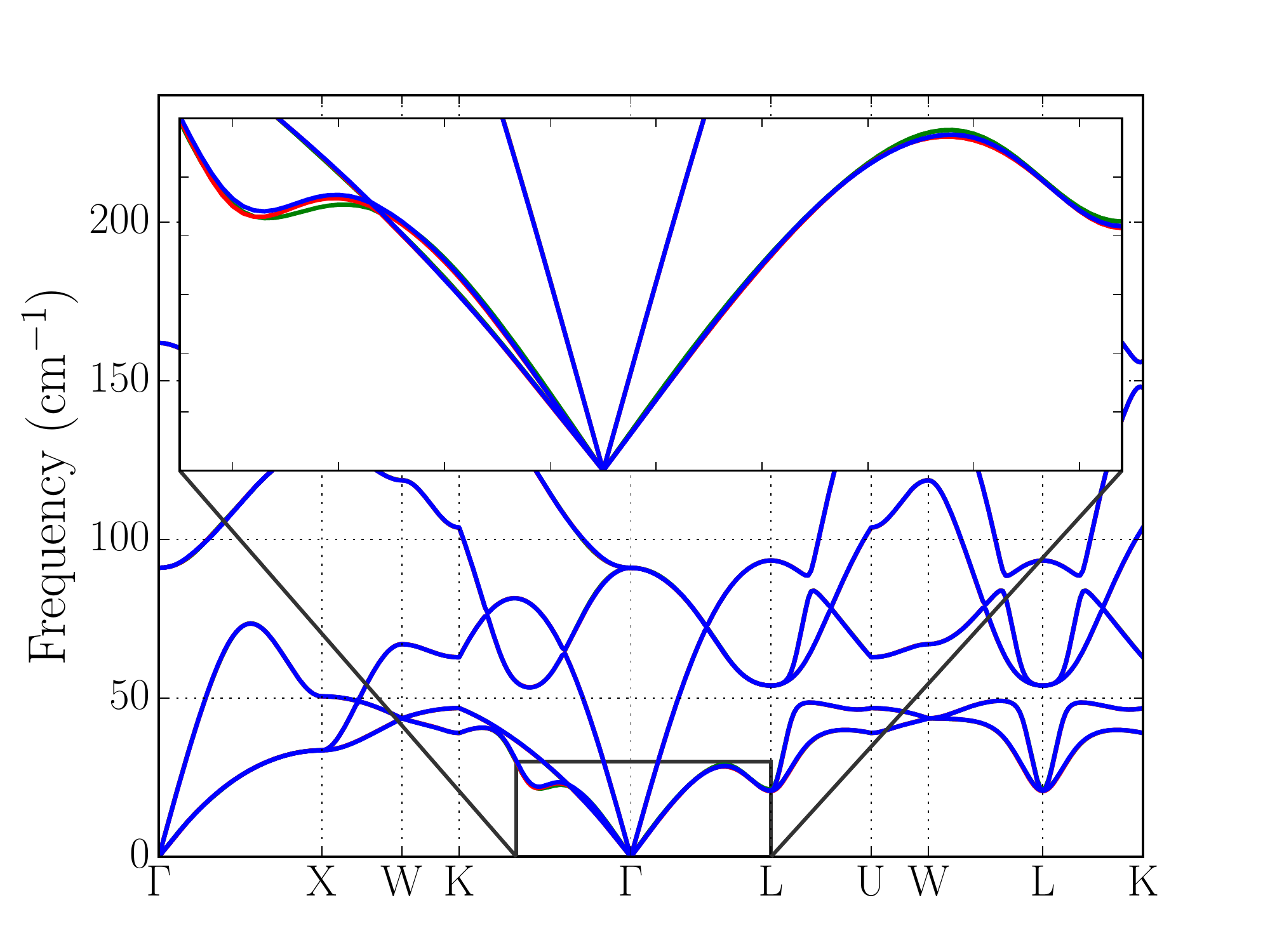}
 \caption{AgCl phonon band structure with equal \textit{k}-point and \textit{q}-point grids $8\times8\times8$ and $10\times10\times10$, $12\times12\times12$ grid for green, red and blue lines, respectively. \label{fig_sm:AgCl_bs}}
\end{center}
\end{figure*}

An additional source of error is coming in the case of AgCl (mp-22922) from the presence of softened modes, as shown in Fig. \ref{fig_sm:AgCl_bs}. Here the results are shown with equivalent \textit{k} and \textit{q} grids, in order to remove the contribution coming from the error discussed above. The dips along the line $\Gamma$-K and at L deepens for tensing strain, but already at this stage they hinder the effectiveness of the Fourier interpolation, resulting in a slower convergence of the phonon band structure. In these cases we consider as unimportant the errors emerging in such a region. If a more accurate analysis is needed it would be more appropriate to focus the study on the point of interest with targeted DFPT calculations.

